\documentclass[preprint,           
               showpacs,            
               preprintnumbers,     
               aps,                 
               prd,                 
               a4paper,             
               superscriptaddress,  
               nofootinbib,         
               ]{revtex4}

\usepackage{graphicx}
\usepackage{amssymb,latexsym}

\newcommand{\beq}{\begin{equation}}
\newcommand{\beqa}{\begin{eqnarray}}
\newcommand{\eeq}{\end{equation}}
\newcommand{\eeqa}{\end{eqnarray}}
\newcommand{\p}{\partial}

\newcommand{\simg}{\gtrsim}
\newcommand{\siml}{\lesssim}
\newcommand{\vp}{\varphi}

\def\NPB#1#2#3{Nucl. Phys. {\bf B#1}, #2 (19#3)}

\def\PRD#1#2#3{Phys. Rev. D {\bf #1}, #2 (19#3)}

\begin{document}
\preprint{gr-qc/0502070
}

\title{Generalized Gravity and a Ghost
}

\author{Takeshi Chiba}%
\affiliation{
Department of Physics, College of Humanities and Sciences, \\
Nihon University, 
Tokyo 156-8550, Japan}
\affiliation{%
Division of Theoretical Astronomy, National Astronomical Observatory of 
Japan, Tokyo 181-8588, Japan }

\date{\today}

\pacs{04.80.Cc; 95.35.+d; 98.80.Es}

\begin{abstract}
We show that generalized gravity 
theories involving the curvature invariants of the Ricci tensor and the 
Riemann tensor as well as the Ricci scalar are equivalent to multi-
scalar-tensor gravities with four-derivatives terms. 
By expanding the action around a vacuum spacetime, the action is reduced 
to that of the Einstein gravity with four-derivative terms, and consequently 
there appears a massive spin-2 ghost in such generalized gravity theories 
in addition to a massive spin-0 field. 
\end{abstract}

\maketitle

\section{Introduction}

The problem of dark energy is the problem of 
$\Omega$: $\Omega=8\pi G\rho_M/3H^2 <1$. Since $\Omega$ can be 
regarded as the ratio of the right-hand-side of the Einstein equation 
(matter) to the left-hand-side of the Einstein equation (curvature=gravity), 
in order to make $\Omega=1$ one requires either (i) introduction of new 
form of matter(energy): dark energy or (ii) modification of gravity in the 
large, so that the total energy density is equal to the critical density, 
which is required by theory (inflation) or by observation (WMAP). 

Recent attempts to modify gravity by introducing a $R^{-1}$ 
term \cite{cdtt,cct} (or more generally a function of the Ricci scalar $F(R)$) 
fall into the latter possibility. However, it is found that 
such a gravity theory is equivalent to the Brans-Dicke theory with 
vanishing Brans-Dicke parameter $\omega$ and a potential term. 
The effective mass 
of the Brans-Dicke scalar field in CDTT model is found to be very 
small ($\sim H_0$). Hence, together with $\omega=0$, 
the solar system experiments exclude such a theory \cite{chiba,Flanagan}.

More recently, a more generalized version of modified gravity theories is 
proposed whose Lagrangian 
is a general function of curvature tensors \cite{cddett} 
(seel also \cite{mb}):
\beq
S=\int \sqrt{-g}d^4x F(R,R_{ab}R^{ab},R_{abcd}R^{abcd}).
\label{action1}
\eeq
The purpose of this note is threefold. We shall firstly show that 
the above action is equivalent 
to multi-scalar-tensor gravity theory with four-derivative terms (Sec.2). 
The equivalent action, however, is still complicated for studying 
observational consequences of the model. 
Therefore, in Sec.3, we expand the action 
around a vacuum spacetime and show that there appear massive spin-2 ghost 
excitations around a vacuum. Applying the previous results to the CDDETT 
model \cite{cddett}, it is found that a spin-2 ghost does appear indeed 
and that the vacuum is unstable. 

\section{Equivalent Action}

Introducing auxiliary scalar fields $\phi_1,\phi_2$ and $\phi_3$, the action 
Eq.(\ref{action1}) is equivalent to 
\beq
S=\int\sqrt{-g}d^4x\Bigl{[}F(\phi_1,\phi_2,\phi_3)+
F_1(R-\phi_1)+F_2(R_{ab}R^{ab}-\phi_2)+
F_3(R_{abcd}R^{abcd}-\phi_3)
\Bigr{]},
\label{action2}
\eeq
where $F_i=\p F/\p\phi_i$. 
It can be found that if the matrix $\p^2 F/\p\phi_j\p\phi_j$ is 
non-degenerate, the equations of motion of $\phi_i~(i=1,2,3)$ give 
$\phi_1=R,\phi_2=R_{ab}R^{ab}$ and $\phi_3=R_{abcd}R^{abcd}$. Putting these 
into Eq.(\ref{action2}) yields the original action Eq.(\ref{action1}). 
Thus the action Eq.(\ref{action1}) is equivalent to the action 
Eq.(\ref{action2}): multi(triple)-scalar-tensor gravity theory with 
four-derivative terms. 

Because of the following identity (the Gauss-Bonnet 
invariant)\cite{dewitt,bo}:
\beqa
&&{\delta\over \delta g^{ab}}\int \sqrt{-g}d^4x(R_{abcd}R^{abcd}-
4R_{ab}R^{ab}+R^2)\nonumber\\
&&=
{\delta\over \delta g^{ab}}\int \sqrt{-g}d^4x\left(C_{abcd}C^{abcd}-
2R_{ab}R^{ab}+{2\over 3}R^2\right)=0,
\label{gb}
\eeqa
where $C_{abcd}$ is the Weyl tensor, Eq.(\ref{action2}) can be rewritten as
\beqa
S&=&\int\sqrt{-g}d^4x\Bigl{[}F(\phi_1,\phi_2,\phi_3)+F_1R+
{1\over 3}(F_2+F_3)R^2+{1\over 2}(F_2+4F_3)C_{abcd}C^{abcd}\nonumber\\
&&-{1\over 2}(F_2+2F_3)\left(C_{abcd}C^{abcd}-2R_{ab}R^{ab}+
{2\over 3}R^2\right)
-\phi_1F_1-\phi_2F_2-\phi_3F_3
\Bigr{]}.
\label{action3}
\eeqa
Introducing the fourth auxiliary field $\phi_4$, the above action is 
further rewritten as
\beqa
S&=&\int\sqrt{-g}d^4x\Bigl{[}F(\phi_1,\phi_2,\phi_3)+
{1\over 3}\left(3F_1+2(F_2+F_3)\phi_4\right)R\nonumber\\
&&+{1\over 2}(F_2+4F_3)C_{abcd}C^{abcd}
-{1\over 2}(F_2+2F_3)\left(C_{abcd}C^{abcd}-2R_{ab}R^{ab}+
{2\over 3}R^2\right)\nonumber\\
&&-\phi_1F_1-\phi_2F_2-\phi_3F_3-{1\over 3}(F_2+F_3)\phi_4^2
\Bigr{]}.
\label{action4}
\eeqa
Hence the gravity theory described by Eq.(\ref{action1}) is equivalent to 
 multi(quadruple)-scalar-tensor gravity whose scalar fields are coupled to 
the curvature terms (Ricci scalar, Weyl squared and the Gauss-Bonnet term). 

\section{Weyl Ghost}

The equivalent action Eq.(\ref{action4}), however, is still complicated 
and is not useful for studying the field content of the theory and 
the experimental constraints on the theory.\footnote{Only recently have 
experimental constraints on single scalar-tensor gravity with the 
Gauss-Bonnet term been studied in \cite{ge}.}  
In order to do so, we limit 
ourselves to studying excitations around a vacuum state (a maximally 
symmetric spacetime with a constant Ricci scalar $R_0$) \cite{how}. 
We introduce trace-free components of the Ricci tensor, 
$R_{ab}=S_{ab}+Rg_{ab}/4$, and  
keep terms up to quadratic order in the curvature, and noting that 
the trace-free 
components are vanishing for maximally symmetric spacetimes, we have
\beqa
S&=&\int d^4x\sqrt{-g} \Bigl{[}F_0+{\p F\over \p R}\Big{|}_0(R-R_0)+
{1\over 2}{\p^2 F\over \p R^2}\Big{|}_0(R-R_0)^2
+{1 \over 2}{\p^2 F\over \p S^2}\Big{|}_0S^2+{1\over 2}
{\p^2 F\over \p C^2}\Big{|}_0C^2
\Bigr{]}\nonumber\\
&=&\int d^4x\sqrt{-g}\Bigl{[} \left(F_0-{\p F\over \p R}\Big{|}_0R_0+
{1\over 2}{\p^2 F\over \p R^2}\Big{|}_0R_0^2\right)
+\left({\p F\over \p R}\Big{|}_0-{\p^2 F\over \p R^2}\Big{|}_0R_0\right)R\\
&&+\left({1\over 2}{\p^2 F\over \p R^2}\Big{|}_0+
{1\over 24}{\p^2 F\over \p S^2}\Big{|}_0\right)R^2
+\left({1\over 2}{\p^2 F\over \p C^2}\Big{|}_0+
{1\over 4}{\p^2 F\over \p S^2}\Big{|}_0\right)C^2
-{1\over 4}{\p^2 F\over \p S^2}\Big{|}_0
\left(C^2-2S^2+{1\over 6}R^2\right)\Bigr{]},
\nonumber
\label{equiv:action}
\eeqa
where subscript $0$ denotes the value evaluated at the maximally symmetric 
spacetime and $C^2=C_{abcd}C^{abcd}$. The final expression is the 
Einstein-Hilbert action with a 
cosmological constant (the first and second term) with a Ricci squared 
term (the third term) and a Weyl squared term (the fourth term) 
and a Gauss-Bonnet term (the fifth term).\footnote{Note that 
$R_{abcd}R^{abcd}-4R_{ab}R^{ab}+R^2=C_{abcd}C^{abcd}-2S_{ab}S^{ab}+R^2/6$.} 
The Gauss-Bonnet term can be dropped classically as a total divergence. 
The action is thus equivalent to
\beqa
S={1\over 16\pi G}\int\sqrt{-g}d^4x\Bigl{(}
R-2\Lambda+{1\over 6m_0^2}R^2-{1\over 2m_2^2}C^2
\Bigr{)},
\label{c^2}
\eeqa
that is, the action of the gravity theory with the four-derivative 
terms ($R^2,C^2$) which has been extensively studied \cite{stelle,mff,how}, 
and it is found that there appear new degrees of freedom: a massive 
spin-0 field (with mass $m_0$) corresponding to $R^2$ 
term (common one for general $F(R)$ 
type gravity) and a massive spin-2 field (with mass $m_2$) 
corresponding to the Weyl squared term. 
Moreover, the massive spin-2 
field is known to have a wrong sign of kinetic term and thus has negative 
energy: a ghost field (the Weyl ghost, also known as 
poltergeist)\cite{stelle}.
\footnote{We note that although the Weyl ghost can be 
eliminated if its mass is infinite (with a fine-tuning), the massive 
scalar field still persists and the problem of $F(R)$ gravity comes back.}

The presence of a ghost may be seen in three ways: i) the direct analysis 
of linearized fields \cite{stelle}, or ii) the analysis of 
propagator \cite{stelle,ns}, or iii) the use of no-go theorem excluding 
consistent cross-couplings between spin-2 fields \cite{ad,bdgh}. We briefly 
mention the last method. Introducing an auxiliary field $\phi$, 
Eq.(\ref{c^2}) can be rewritten as \cite{how}
\beqa
S={1\over 16\pi G}\int d^4x\sqrt{-g}\left[ \left(1+{\phi\over 3m_0^2}R\right)-
{\phi^2\over 6m_0^2}-{1\over 2m_2^2}C^2
\right].
\label{c^2:2}
\eeqa
If we perform the conformal transformation, 
${\tilde g}_{ab}=(1+\phi/3m_0^2)g_{ab}=e^{\vp}g_{ab}$, then Eq.(\ref{c^2:2}) 
is further rewritten as (recall that Weyl squared term is conformally 
invariant) 
\beqa
S={1\over 16\pi G}\int d^4x\sqrt{-{\tilde g}}\left[ {\tilde R}-
{3\over 2}({\tilde \nabla} \vp)^2-{3m_0^2\over 2}(1-e^{-\vp})^2-
{1\over 2m_2^2}{\tilde C}^2
\right].
\label{c^2:3}
\eeqa
Then introducing another auxiliary field $\pi_{ab}$ and 
 using Eq.(\ref{gb}), Eq.(\ref{c^2:3}) is equivalent to
\beqa
S={1\over 16\pi G}\int d^4x\sqrt{-{\tilde g}}\left[ {\tilde R}-
{3\over 2}({\tilde \nabla} \vp)^2-{3m_0^2\over 2}(1-e^{-\vp})^2-
{\tilde G}_{ab}{\tilde \pi}^{ab}+{1\over 4}m_2^2
({\tilde \pi}_{ab}{\tilde \pi}^{ab}-
{\tilde \pi}^2)
\right].
\label{c^2:4}
\eeqa
{}From the equation of motion of ${\tilde \pi}_{ab}$, 
$2{\tilde G}_{ab}=m_2^2({\tilde \pi}_{ab}-{\tilde \pi}{\tilde g}_{ab})$.   
Further it is found from the equation of motion of metric that 
${\tilde \pi}_{ab}$ field satisfies a generalized transverse traceless 
condition: 
spin-2 field \cite{how}. Thus, the fourth term in Eq.(\ref{c^2:4}) 
represents the cross interaction term between spin-2 fields 
(${\tilde g}_{ab}$ and ${\tilde \pi}_{ab}$) which is excluded by 
the no-go theorem \cite{ad,bdgh}. 
According to the no-go theorem, there is no consistent (ghost-free) 
coupling between spin-2 fields: such a coupling necessarily 
yields ghosts.

\section{Application}
For the case of CDDETT model \cite{cddett}
\beq
S={1\over 16\pi G}\int d^4x\sqrt{-g}\left(R-{\mu^{4n+2}\over 
(aR^2+bR_{ab}R^{ab}+cR_{abcd}R^{abcd})^n}\right),
\label{cddett}
\eeq
by expanding the action around a maximally symmetric spacetime with Ricci 
scalar $R=R_0$ up to the quadratic order in the curvature, we obtain
\beqa 
S&=&{1\over 16\pi G}\int d^4x\sqrt{-g}\Bigl{[}
-(n+1)(2n+1)\lambda +\left(1+{4n(n+1)\lambda\over R_0}\right)R\nonumber\\
&&-{n\lambda(6(2n+1)a+(3n+1)b+2nc)\over 6d R_0^2}R^2+
{n\lambda(b+4c)\over 2d R_0^2}C_{abcd}C^{abcd}
\Bigr{]},
\eeqa
where $d=a+b/4+c/6,\lambda=\mu^{4n+2}/(dR_0^2)^n$. 
Comparing with Eq.(\ref{equiv:action}), we find that the effective 
mass squared, for both 
a scalar field of freedom and a spin-2 field of freedom, is negative 
(tachyonic) as long as $a,b,c$ and $n$ are positive, 
which implies that the vacuum is unstable. Note that this conclusion is 
consistent with the analysis in \cite{cddett}, although our analysis 
is not limited to homogeneous and isotropic cosmological models. 
In any case, the presence of the Weyl squared term 
implies the presence of a ghost field. 

We may estimate the order-of-magnitude of the effective mass squared,  
$m_0^2$ and $m_2^2$, for the CDDETT model as an alternative to dark energy. 
In order that the 
``correction term'' (the second term) in Eq.(\ref{cddett}) plays the role of 
dark energy  (presently dominant component of the universe), 
it should be comparable to (or larger than) the first term 
(the Einstein-Hilbert term). Thus, for $R\simeq H_0^2$, 
$\mu^{4n+2}/H_0^{4n}\simg H_0^2$, therefore $\mu \simg H_0$. 
Here we have assumed that the coefficients $a,b$ and $c$ are order unity. 
Since $\lambda \simg H_0^2$, we thus obtain 
\beq
|m_0^2|\simeq |m_2^2| \simeq R_0^2/\lambda\siml H_0^2, 
\eeq
which are very light, as one 
might have expected, and are effectively massless as far as the local 
experiments (like solar system experiments) are concerned. 

A linearized static spherically symmetric solution for Eq.(\ref{c^2}) is 
obtained in \cite{stelle} and is given by 
\beqa
ds^2&=&-(1+2U)dt^2+(1+2V)dr^2+r^2d\Omega^2\\
&&U=-{GM\over r}-{GM\over 3}{e^{-m_0r}\over r}+
{4GM\over 3}{e^{-m_2r}\over r}\nonumber\\
&&V={GM\over r}-{GM\over 3}{e^{-m_0r}\over r}-{4GM\over 3}{e^{-m_2r}\over r}
-{GM\over 3}m_0e^{-m_0r}-{2GM\over 3}m_2e^{-m_2r}.\nonumber
\eeqa
For vanishing $m_0$ and $m_2$, we have $U=0$ and $V=-2GM/r$. Hence 
the correct Newtonian limit is not reproduced and the 
PPN (parameterized post-Newtonian) parameter $\gamma=-V/U$ is undefined. 
(We also note that if $m_0=0$ and $m_2$ is very large, then $\gamma=1/2$, 
which corresponds to the Brans-Dicke parameter $\omega=0$ being consistent 
with \cite{chiba}.)

Hence the gravity model of this type is much more problematic than $F(R)$ type 
gravity: it contains new gravitational degrees of freedom in addition to 
massless spin-2 field: an almost massless scalar field and a massive spin-2 
ghost (a state with negative energy). It is not viable gravity 
theory from both phenomenological (solar system experiments) and 
theoretical (consistency) point of view.\footnote{Similar observations 
are made in \cite{ns} by analysing propagator directly along the line 
of \cite{stelle}.}

\section{Summary}

We have shown that the generalized gravity of 
$F(R,R_{ab}R^{ab},R_{abcd}R^{abcd})$ 
type is equivalent to multi-scalar-tensor gravity theory with 
four-derivative terms. It would be interesting to investigate 
the weak field gravity in such gravity theories. 
We have also shown that the degrees of freedom of the generalized gravity of 
$F(R,R_{ab}R^{ab},R_{abcd}R^{abcd})$ type is equivalent 
to that of $R^2+C^2$ gravity. It contains a massive scalar field 
(Brans-Dicke type scalar) and a massive spin-2 field. 
The massive spin-2 field has negative energy and is a ghost. 
In quantum theory, negative energy must be traded for quantum states 
with positive energy but this time with negative norm, thus losing unitarity. 
If such modifications became important recently, the scalar field and massive 
spin-2 field are generically very light and mediate a gravity force of long 
range. Hence such a theory cannot be a viable gravity 
theory from both phenomenological (solar system experiments) and 
theoretical (consistency) point of view. 
As a side remark, we note that our analysis makes use of a topological 
identity (Gauss-Bonnet identity) which is valid only in four spacetime 
dimensionsand hence does not preclude the presence of higher curvature 
terms in higher dimensional theory. 

\begin{acknowledgments}
We would like to thank Prof. K. Maeda at the early stage of this work. 
This work was supported in part by a Grant-in-Aid for Scientific 
Research (No.15740152) from the Japan Society for the Promotion of
Science. 
 \end{acknowledgments}


\end{document}